\documentclass[aps, prx, reprint, longbibliography]{revtex4-2}
\usepackage{xr-hyper, hyperref, graphicx, amsmath, amssymb, xcolor,lineno}
\graphicspath{{figures/}}

\begin{document}
\title{Universal reconstructive polarimetry with graphene-metal infrared photodetectors
}

\author{Valentin Semkin$^{1}$}
\email[]{semkin.va@phystech.edu}

\author{Kirill Kapralov$^{1,2}$}

\author{Ilya Mazurenko$^{1,3}$}

\author{Mikhail Kashchenko$^{1,2,3}$}

\author{Alexander Morozov$^{3}$}

\author{Yakov Matyushkin$^{1}$}

\author{Dmitry Mylnikov$^{1}$}

\author{Denis Bandurin$^{4}$}

\author{Li Lin$^{5}$}

\author{Alexey Bocharov$^{1,2}$}

\author{Dmitry Svintsov$^{1,2}$}
\email[]{svintcov.da@mipt.ru}

\affiliation{$^{1}$Moscow Institute of Physics and Technology, Dolgoprudny 141700, Russia}
\affiliation{$^{2}$Joint-Stock Company ''Skanda Rus'', Krasnogorsk 143403, Russia}
\affiliation{$^{3}$Programmable Functional Materials Lab, Center for Neurophysics and Neuromorphic Technologies, Moscow 127495, Russia }
\affiliation{$^{4}$Department of Materials Science and Engineering, National University of Singapore, 117575, Singapore}
\affiliation{$^{5}$School of Materials Science and Engineering, Peking University, Beijing, P. R. China}

\begin{abstract}
Recent advent of smart photodetectors, where in-situ tuning of responsivity enables the reconstruction of light intensity, polarization and spectrum by a single device, has revolutionized the field of optoelectronics. So far, most such reconstructive detectors were realized with non-scalable technology of van der Waals stacking. Here, we demonstrate the infrared reconstructive polarimetry with photodetectors based on conventional gated graphene-metal junctions. The reconstruction exploits the gate tuning of polarization contrast, which enables the determination of both infrared power and polarization angle from photovoltage measurements at two different gate voltages. The physics enabling the  polarimetry lies in polarization-dependent shift of the electron hot spot near the contact, and the gate tuning of photosensitive barrier width. We further show the universality of polarization reconstruction, i.e. its feasibility with different geometries of the junction, and with graphene of different quality, from boron-nitride encapsulated flakes to the scalable chemical vapor deposited films. 


\end{abstract}
\maketitle

{\it Introduction}. Rapid progress in automotive transport, robotics, and internet-of-things demands fast and in-depth information extraction from optical scenes~\cite{Deep_otpical_sensing}. Stimulated by these needs, the concepts of smart optical sensors have emerged and evolved~\cite{Computational_review_1,Computational_review_2}. Contrary to conventional sensors perceiving light intensity only, the smart sensors possess extra functionality including real-time processing, classification and encoding of images~\cite{Mennel_encoder,Classifier_Wu}, event detection~\cite{EventSensor_2}, reconstruction of spectrum~\cite{CompSpectr_review,Yuan2021,Yoon2022,Yang2019,Intelligent_infrared_TBG} and polarization of light~\cite{Deng_metasurface_array,Wei_calibration_free,AoLP_DoLP_van_der_waals,Bullock_polarization_resolving,Full_Stokes_graphene}. A broad class of smart single-sensor spectrometers and polarimeters is represented by miniaturized multiple photodetectors, each having its own spectral~\cite{Yang2019,MS_multidimensional_PD} and polarization~\cite{Full_Stokes_graphene,Deng_metasurface_array,Ganichev_ellipticity2} response. A more recent and innovative approach with stronger potential for miniaturization is represented by tunable smart sensors. Here, real-time variations of control parameter (typically, electrical voltage) change the wavelength-~\cite{Yuan2021,Yoon2022,Muravev2012_detector_spectrometer} and polarization-dependent~\cite{Bullock_polarization_resolving,Ferroelectric_P_resolving} responsivity. Eventually, measurements of photosignal at different control parameters enable computational reconstruction of light spectrum, polarization and intensity with a single sensor.

For unknown historical reasons, the first single-sensor spectrometers and polarimeters were demonstrated with van der Waals flakes and heterostructures of two-dimensional materials~\cite{Yuan2021,Yoon2022,Intelligent_infrared_TBG}. Despite rich functionalities, such technology is hardly scalable to high-resolution cameras. Only recently have single-sensor spectrometers been fabricated with III-V-based graded-gap photodiodes~\cite{III_V_cascade_diode_spectrometer} and organic semiconductors~\cite{Organic_semicond_spectrometr}. The existing demonstrations of reconstructive polarimeters are still limited by van der Waals flakes. Absence of scalable material platforms for reconstructive polarimetry obstructs its numerous applications, including high-contrast imaging under limited atmospheric visibility~\cite{Polarimetric_imaging}, material inspection and defect detection~\cite{Polarimetric_imaging_crystals}, chemical and isomeric analysis~\cite{Polarimetry_chiral} and communications with polarization multiplexing~\cite{Polarization_multiplexing}.

Here, we implement the polarimetry of mid-infrared linearly polarized light with conventional detectors based on gated graphene and asymmetric metal electrodes~\cite{Semkin2024,Asymmetric_detectors_2,Wei_geometric_maximizing_BPVE}. Their operation is based on non-compensating zero-bias photovoltages at metal-graphene Schottky junctions. The junction photovoltage emerges via light-induced carrier heating followed by thermoelectric effect~\cite{Tielrooij-JPCM,Tielrooij-NNano}. The technology of such detectors is well-established both for high-quality hBN-encapsulated samples~\cite{Castilla_plasmonic_antenna} and for scalable chemical vapor deposited films~\cite{Semkin2024,Asymmetric_detectors_2,Wei_geometric_maximizing_BPVE}. It is attractive due to the absence of complex chemical doping stages. The physical origin of polarization-resolving action lies in (1) the dependence of hot carrier temperature profile on light polarization via the lightning-rod effect~\cite{Nikulin2021} (2) the tuning of photosensitive Schottky barrier width by the gate voltage~\cite{Khomyakov_screening}. Roughly speaking, tuning of gate voltage enables spatial 'sampling' of hot carriers generated by with light with different polarization states. The polarization-resolving action is specific to the infrared, where the width of photosensitive barrier is comparable with the scale of local field enhancement near the metal contact. Though the gate tuning of polarization contrast in graphene-metal detectors was reported before~\cite{Semkin_gate_controlled,Wei_configurbale_polarity}, the effect was too weak to enable reconstructive polarimetry.

\begin{figure*}[ht]
    \includegraphics[width=1.0\linewidth]{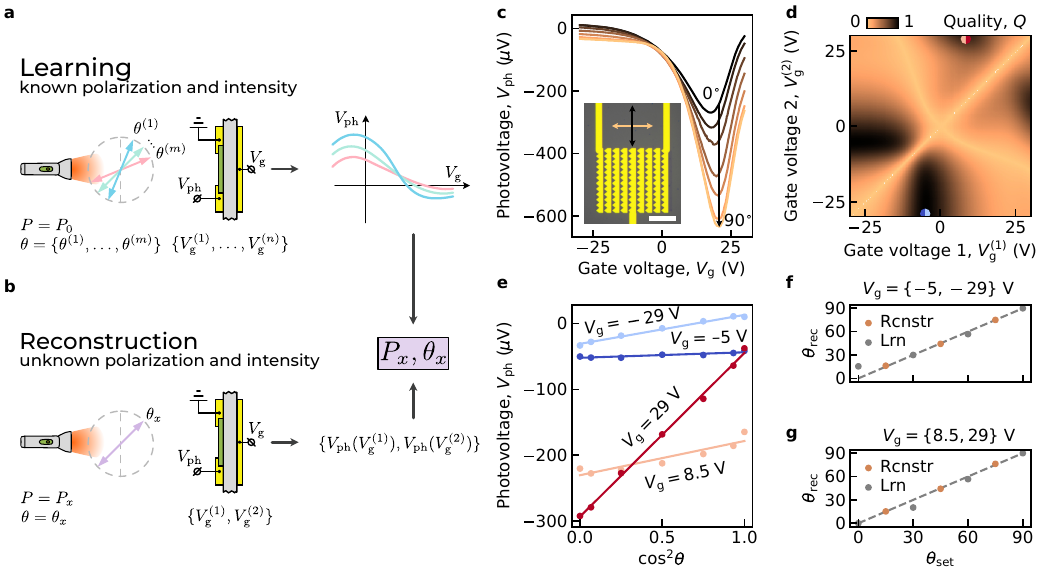}
    \caption{{\bf Principle and demonstration of reconstructive polarimetry with gate-controlled graphene detector.} The principle of reconstructive polarimetry is based on calibration of device responsivity $R$ vs polarization angle $\theta$ and control gate voltage $V_{\rm g}$. The procedure is shown in panel (a). The reconstruction, shown in (b), is based on readout of photovoltage $V_{\rm ph}$ at two subsequent control voltages $V_{\rm g}^{(1),(2)}$. This is generally sufficient to determine two radiation characteristics: power $P_x$ and polarization angle $\theta_x$. (c) Gate-and polarization-dependent photovoltage for graphene detector with geometrically patterned contacts shown in the inset. Scale bar is 20 $\mu$m. (d) Computed quality of polarization resolution $Q$ as a function of control voltage pair used for reconstruction. Optimal working points are marked by color circles. (e) Polarization-dependent photovoltage of metal-graphene detector showing large variations of polarization contrast with gate voltage. (f,g) Demonstration of polarization reconstruction for operation at selected combinations of gate voltage $V_{\rm g} = \{-5,\,-29\}$ V (f) and $V_{\rm g} = \{8.5,\,29\}$ V (g). Gray circles represent the angles used at the calibration stage (marked as 'Lrn'), orange circles represent the angles subject to computational reconstruction (marked as 'Rcnstr').}
    \label{fig1}
\end{figure*}

The paper is organized as follows. We first present model foundations of reconstructive polarimetry and derive a quantitative merit of detector's capability to resolve the linear polarization. We proceed to the demonstration of polarization angle reconstruction using graphene detectors with geometrically patterned metal contacts. After that, we show that numerous graphene-metal detector architectures are capable of polarization resolution. We conclude the presentation by developing the minimal physical model of polarization-resolving response at a linear metal-graphene junction.

{\it Fundamentals of the reconstructive polarimetry}. The functional dependences of detector photoresponse $V_{\rm ph}$ on polarization angle $\theta$ are constrained to certain trigonometric functions. This fact simplifies greatly the procedure of reconstructive polarimetry. More precisely, the dc photovoltage of a detector is a quadratic function of optical electric field components $E_\alpha E_\beta^*$, the latter combinations known as Stokes parameters. In the absence of chirality and for known principal axes of the detector, only two Stokes parameters are relevant. One can select the total optical power $P$ and the power associated with a certain symmetry axis $P \cos^2\theta$ for such parameters. The photovoltage appears as $V_{\rm ph} = R_{\rm is} P + R_{\rm an} P\cos^2\theta$, where the quantities $R_{\rm is}$ and $R_{\rm an}$ can be called isotropic and anisotropic responsivities.

The reconstructive polarimetry, i.e. simultaneous determination of optical power and polarization angle, becomes possible once isotropic and anisotropic responisivities are configurable by an external voltage. Such voltage $V_{\rm g}$ is hereby applied to the global gate, which is convenient in the technology of graphene detectors formed on doped Si/SiO$_2$ substrates. Denoting the array of gate voltages with ${\bf V}_{\rm g} =\{V_{\rm g}^{(1)},...V_{\rm g}^{(n)}\}$, $n\ge 2$, and the respective arrays of tunable responsivities as ${\mathbf R}_{\rm is}$ and ${\mathbf R}_{\rm an}$, we arrive at a basic equation of reconstructive polarimetry:
\begin{equation}
\label{eq-main_vec}
    {\mathbf V}_{\rm ph} = {\mathbf R}_{\rm is} P +{\mathbf R}_{\rm an} P \cos^2\theta.
\end{equation}
With these physical prerequisites, the reconstruction procedure is reached in two stages. At the learning stage, Fig.~\ref{fig1} a, the detector response is recorded at variable polarization angle $\theta =\{\theta^{(1)},...\theta^{(m)}\}$ and set of gate voltages. The $\cos^2\theta$-polarization dependence at each gate is verified, and calibration arrays ${\mathbf R}_{\rm is}$ and ${\mathbf R}_{\rm an}$ are formed upon fitting. At the reconstruction stage, the sensor is illuminated with light of unknown power $P_x$ and polarization $\theta_x$, Fig.~\ref{fig1} b. One measures the photoresponse array ${\bf V}_{\rm ph}$ at several gates $\{V_{\rm g}^{(1)},...V_{\rm g}^{(k)}\}$~\footnote{The array of gate voltages at the calibration stage is generally larger than that at the measurement stage, i.e. $N \ge k \ge 2$}. Both power and polarization angle are computed from basic equation (\ref{eq-main_vec}), using the measured ${\bf V}_{\rm ph}$ and calibrated ${\mathbf R}_{\rm is/an}$ as input.

Photovoltage measurements at two control gate voltages $\{V^{(1)}_{\rm g},V^{(2)}_{\rm g}\}$ are, in principle, sufficient for reconstructive polarimetry with model (\ref{eq-main_vec}). Larger input with $k>2$ gate voltages increases the accuracy, the overdetermined system (\ref{eq-main_vec}) is thereby solved with least-squares method. 

The reconstruction procedure described above relies on a subtle physical property of the polarization response: the gate- and polarization dependences of photovoltage should be entangled. In other words, application of gate voltage should change the very pattern of polarization sensitivity, not merely the magnitude of the photoresponse. In prior detection experiments, this property was not observed. The polarization sensitivity $f(\theta)$ was largely governed by external antennas and contacts~\cite{Tielrooij-JPCM,Badioli2014,Castilla_plasmonic_antenna} or crystal lattice~\cite{PdSe2_anisotropic_detection}, being insensitive to the gate and carrier density. In notations of Eq.~(\ref{eq-main_vec}), the situation corresponded to parallel vectors ${\mathbf R}_{\rm is}$ and ${\mathbf R}_{\rm an}$, which made simultaneous determination of $\theta$ and $P$ impossible.

We're now ready to derive the merit of gate-dependent isotropic and anisotropic responses which reflects the ability of the sensor to resolve the polarization. The purpose of this derivation is twofold. First, the merit would show whether the particular sensor can at all disentangle the polarization of light. Second, it would allow a smart selection of working points, i.e. the small subset of gate voltages used for actual polarimetric imaging.
To achieve this, we introduce the reconstruction quality $Q(\mathbf{V}_{\rm g})$ measuring the propagation of error from photovoltage $\delta {\bf V}^2_{\rm ph}$ to the reconstructed $\delta \cos^2\theta$ and $\delta P$:
\begin{equation}
    \frac{\delta {\bf V}^2_{\rm ph}}{\mathbf{V}^2_{\rm ph}} = \frac{Q^2(\mathbf{V_{\rm g}})}{2} \left[(\delta \cos^2\theta)^2+\left(\frac{\delta P}{P}\right)^2\right].
\end{equation}
The variational and statistical analysis of the polarization-dependent responsivity~(\ref{eq-main_vec}) results in the following quality merit~(see Supplementary section I for derivation):
\begin{equation}
\label{eq-quality}
    Q^2({\bf V}_{\rm g}) =   \frac{2\sin^2\phi}{\left(r+\frac{3}{8r}+\cos\phi\right)\left(r+\frac{11}{8r}+\cos\phi\right)},
\end{equation}
where $r^2 = {\bf R}^2_{\rm is}/{\bf R}^2_{\rm an}$ and $\cos\phi = ({\bf R}_{\rm is},{\bf R}_{\rm an})/(|{\bf R}_{\rm is}||{\bf R}_{\rm an}|)$. The quality merit varies between zero and unity, where unity corresponds to the best available polarimetry, and zero corresponds to the impossibility to determine $\theta$ from given measurements. The polarization reconstruction quality (\ref{eq-quality}) is maximized for $r^2=5/8$ and $\cos\phi = -\sqrt{2/5}$. The above result quantitatively explains how much tunable should be the responsivity to provide accurate polarization reconstruction. 

Mapping of quality $Q(V_{\rm g}^{(1)},V_{\rm g}^{(2)})$ and its maximization with respect to the pairs of gate voltages should be preferably introduced between the learning and reconstruction stages. In actual sensors, the absolute maximum $Q_{\max} = 1$ may be not reached, still, we shall show that the angle reconstruction quality is readily close to unity for several sensor designs. 


{\it Experimental demonstration of polarimetry with graphene-metal infrared detector}. 
Our first polarization-resolving device is a graphene channel photodetector with a metasurface of wedge-structured metallic electrodes atop of it, and doped silicon gate beneath 285 nm of silicon dioxide~\cite{Semkin2024}. The radiation-sensitive graphene layer is grown with scalable chemical vapor deposition (CVD) method and wet-transferred onto the substrate. The photograph of the structure is shown in the inset of Fig.~\ref{fig1} c. The overall device size is $A = 45 \times 45$ $\mu$m$^2$, it is comparable with the size of focused laser beam $\sigma = 20$ $\mu$m.

Inversion asymmetry of the metal pattern enables zero-bias photocurrent. It mainly occurs due to the photo-thermoelectric effect at wedge-shaped metal-graphene junctions. The gate- and polarization dependences of the photocurrent, recorded upon illumination with quantum cascade laser with $\lambda_0 = 8.6$ $\mu$m, are shown in Fig.~\ref{fig1}c. As the radiation ${\bf E}$-field is directed along the wedge axis ($\theta = 90^\circ$), graphene electrons in the vicinity of vertex are overheated, and subsequently drift toward 'cold' non-patterned electrode, which results in large photovoltage. For orthogonal polarization, the field enhancement by the wedge is absent, which results in reduced photovoltage. Such a large polarization ratio (PR) holds for almost all gate voltages except the range of $V_{\rm g}\approx -5...5$ V, where the contrast disappears. 

Retaining the discussion of physics governing the vanishing contrast to the last section, we focus on using the effect for polarization resolution. In Fig.~\ref{fig1} e, we verify the linear dependence of photovoltage on $\cos^2\theta$, and extract the learning functions $R_{\rm is}(V_{\rm g})$ and $R_{\rm an}(V_{\rm g})$. 

It is remarkable that both isotropic and anisotropic parts of responsivity of  Fig.~\ref{fig1} (e) are largely variable with gate voltage. The absolute change in photovoltage upon 90$^\circ$ polarization rotation $\Delta V_{\rm ph}$ changes from $\sim 250$ $\mu$V at $V_{\rm g} = +29$ V to $<10$ $\mu$V at $V_{\rm g} = -5$ V. This corresponds to almost complete disappearance of polarization contrast upon gating. At the same time, the photovoltage remains quite large at the polarization insensitivity point, $V_{\rm ph}\sim 100$ $\mu$V, which is $\sim 6$ times below the maximum photovoltage. It implies that both polarization-sensitive and polarization-insensitive photovoltages are readily measurable above the noise level.

The feasibility of polarization reconstruction is governed not by polarization contrast, but rather by the newly introduced quality function (\ref{eq-quality}). We plot it in Fig.~\ref{fig1} d versus two gate voltages. The quality is maximized for the two pairs $\{ V_{\rm g}^{(1)},V_{\rm g}^{(2)} \} = \{-5\, {\rm V},-29 \, {\rm V}\}$ and $\{ V_{\rm g}^{(1)},V_{\rm g}^{(2)} \} = \{8.5\, {\rm V},29 \, {\rm V}\}$. Notably, none of the gate voltage combinations corresponds to the maximum and minimum polarization ratios.

The resulting polarimetry for the two optimal pairs of gate voltages ($-$5 V and $-$29  V in Fig.~\ref{fig1} f, and 8.5 V and 29  V in Fig.~\ref{fig1} g) is demonstrated in Figs.~\ref{fig1} f and g. We use four values of $\theta_{\rm lrn}$ (marked with gray in figures) to learn $R_{\rm is}(V_{\rm g})$ and $R_{\rm an}(V_{\rm g})$ at these particular gate voltages. After that, we set the polarizer to the positions $\theta_{\rm set}$ different from those used at the learning stage. We measure the pairs of photovoltages ${\bf V}_{\rm ph} = \{V_{\rm g}^{(1)},V_{\rm g}^{(2)}\}$, and reconstruct the angles by solving the quasi-linear system (\ref{eq-main_vec}). The reconstructed angles $\theta_{\rm rcnstr}$ are shown in Figs.~\ref{fig1} f and g with orange dots. The angles determined by the reconstruction algorithm $\theta_{\rm rcnstr}$ fall very close to the values $\theta_{\rm set}$. The relative error $|\theta_{\rm set}-\theta_{\rm rcnstr}|/\theta_{\rm set}$ is less than 8.2~\%, which implies very good reconstruction quality.

\begin{figure*}[ht]
    \includegraphics[width=1\linewidth]{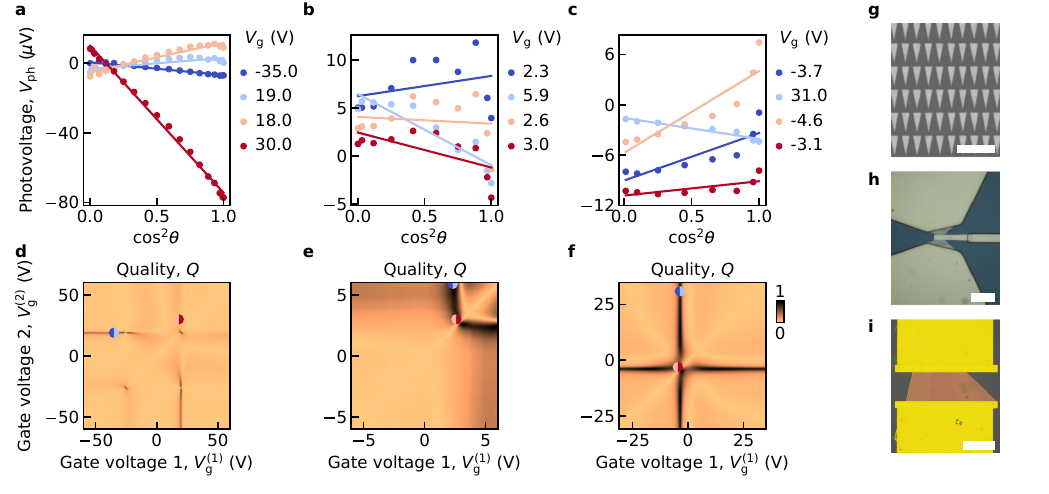}
    \caption{{\bf Universality of the reconstructive polarimetry with dissimilar graphene photodetectors.} Panels (a-c) show the experimentally measured polarization-dependent photovoltages at different gate voltages for three different detector structures. All devices feature gate-tunable polarization contrast. Micro-photographs of the detectors are shown in panels (g-i): (g) hBN-encapsulated graphene photodetector with different source and drain widths (h) photodetector with non-parallel source and drain contacts and a top gate (i) photodetector with inversion-asymmetric metallic metasurface deposited directly atop graphene. Panels (d-f) show the computed quality of polarization reconstruction for the three respective detectors. Scale bars are 2 $\mu$m (g), 20 $\mu$m (h) and 50 $\mu$m (i).}
    \label{fig2}
\end{figure*}

{\it Universality of reconstructive polarimetry with graphene-metal detectors}. Surprisingly, the polarization-resolving response was identified not only in a single wedged detector architecture, shown in Fig. \ref{fig1} c. Numerous detectors based on graphene-metal asymmetric contacts appeared to have 'entangled' gate- and polarization dependent response in the mid-infrared. In other words, application of gate voltage resulted in largely variable polarization sensitivity. This property enables the reconstruction of the linear polarization angle. The ubiquitous character of polarization-resolving response is illustrated in Fig.~\ref{fig2}.

We start by modifying the structure of the metasurface atop the graphene layer, and disconnect the individual wedges by preserving the non-centrosymmetric structure~\cite{Wei_geometric_maximizing_BPVE}. The resulting device is shown in Fig.~\ref{fig2} (i). Its gate- and polarization-dependent photoresponse, recorded with the same $\lambda_0 = 8.6$ $\mu$m laser, is shown in Fig.~\ref{fig2} (c). The photovoltage changes sign upon the $90^\circ$ rotation of the polarization, which can be explained by different positions of local hot spots. When the ${\bf E}$-field is directed along the triangle axis, the local intensity hot spot is positioned at the apex. For $90^\circ$ rotated polarization, the hot spot is at the triangle base, resulting in inversion of carrier temperature gradient and hence, photovoltage. 

The slope and magnitude of $V_{\rm ph}(\cos^2\theta)$--dependences are highly controllable by the gate voltage in the 'disconnected metasurface' sensor, similarly to our first example. Quantitative analysis of polarization resolution merit $Q$, however, shows that the reconstruction quality is hereby pretty low, $Q\lesssim 0.1$, Fig.~\ref{fig2} (f). Still, we managed to demonstrate the resolution of polarization angle with this structure by recording $V_{\rm ph}$ at a large set of gate voltage points $k = 30$. The results are shown in Supplementary section II. Remarkably, such resolution was achieved at various levels of incident power.

After these two examples, it is tempting to reveal whether polarization-resolving response is specific to graphene-metal metasurfaces, or it can persist in simpler detector structures. To answer the question, we record the mid-IR response of CVD-graphene channel detector with a top gate and two source and drain contacts connected to an asymmetric antenna. The device is shown in Fig.~\ref{fig2} (h). In prior experiments, it was used for studies of chiral terahertz photoresponse enabled by antenna asymmetry~\cite{Matyushkin2020}. This asymmetry also provides non-zero photovoltage in the infrared: at any ${\bf E}$-field orientation deviating from symmetry axis, the local light intensities at the source and drain are different due to the polarization dependence of the lightning-rod effect. The dependences of zero-bias response $V_{\rm ph}$ and $\cos^2\theta$ at various $V_{\rm g}$ for that device are shown in Fig.~\ref{fig2} (b). The linear fits work poorly hereby, as compared to the metasurface-based sensors, due to the small size of the radiation spot compared to the size of antenna~\footnote{More precisely, the linear relation $V_{\rm ph} = R_{\rm is} P + R_{\rm an} P \cos^2\theta$ holds only for uniform illumination. For non-uniform illumination, $V_{\rm ph}$ depends not only on Stokes parameters, but also on their gradients. If the light beam is uniform at the scale of individual photosensitive cell (as it was in the case of metasurface sensors), the gradient terms are irrelevant.}. Nevertheless, the extracted slopes and offsets of polarization-dependent photovoltage are controlled by the gate, which results in elevated reconstruction quality $Q$, Fig.~\ref{fig2} (e). This result shows that the role of metasurface is not determinative for the presence of polarization-resolving action.

We finally show the persistence of polarization-resolving action for the {\it simplest} infrared detectors based on metal-graphene contacts. To this end, we record the photovoltage of graphene channel with parallel source and drain contacts of dissimilar width, shown in Fig.~\ref{fig2} g. The channel is made of hBN-encapsulated graphene and has length of $50$ $\mu$m and width varying from $50$ $\mu$m to 40 $\mu$m. The infrared photoresponse, shown in Fig.~\ref{fig2} a, generally peaks for light ${\bf E}$-field orthogonal to the metal leads. However, in a range of gate voltages near charge neutrality, the polarization contrast becomes low or even changes sign. The anticipated reconstruction quality for the latter structure is shown in Fig.~\ref{fig2} (d) and peaks near the charge neutrality.

{\it Physical origin of the polarization-resolving action}. We have evidenced the gate-control of polarization sensitivity in multiple graphene-metal devices. While the metasurface highlights the polarization-resolving response via singular enhancement of absorbance at the edges, the basic effect of entangled gate-and polarization sensitivity should be present already at the linear graphene-metal edge. Below we suggest a theory of that phenomenon.

\begin{figure*}[ht]
    \includegraphics[width=1\linewidth]{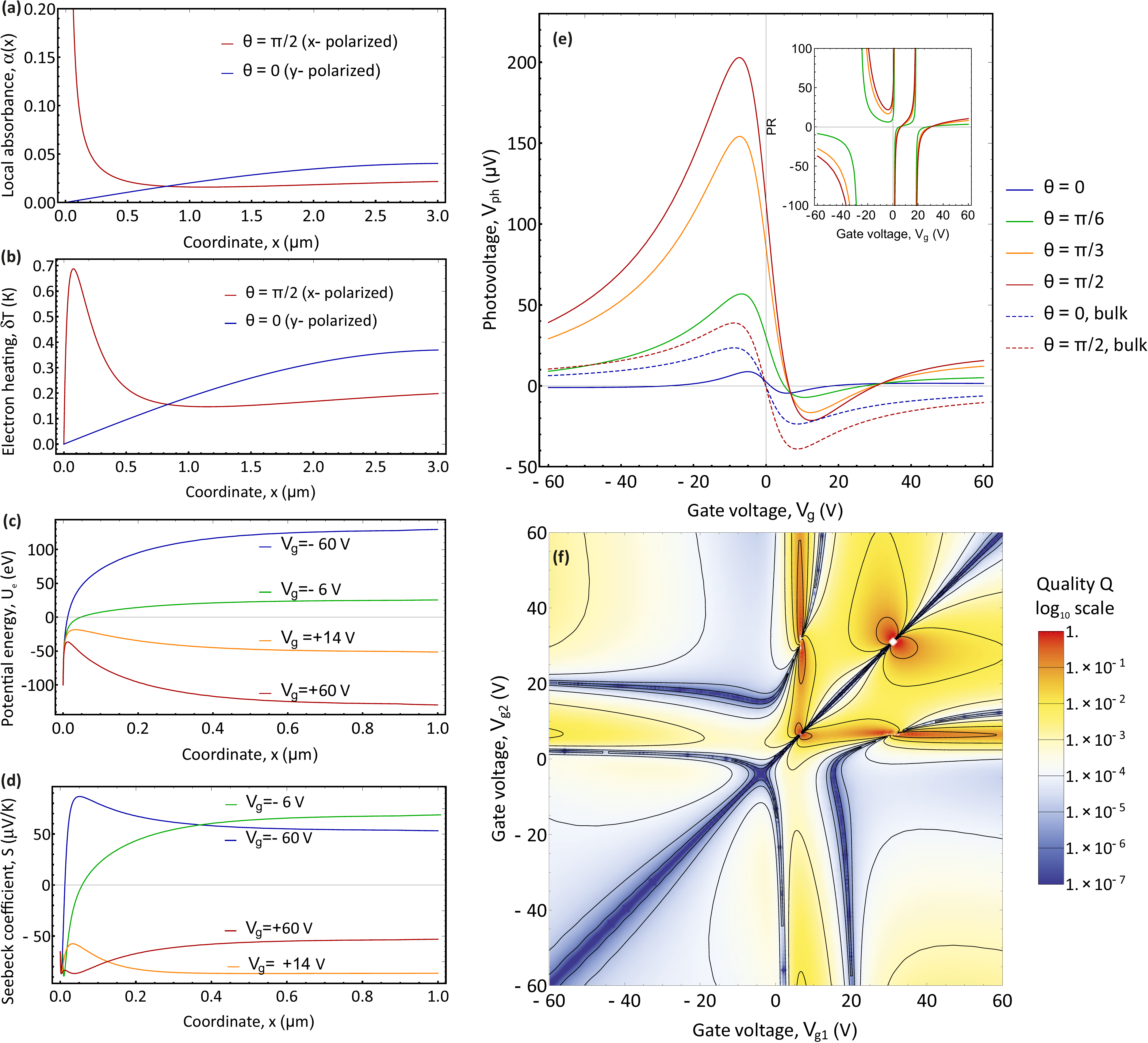}
    \caption{{\bf Theoretical modeling of polarization-resolving detection at a metal-graphene interface.} {\bf (a-d)} Microscopic distributions of various physical quantities at metal-graphene junctions upon mid-IR illumination. {\bf (a)} - local absorbance {\bf (b)} - light-induced change in electron temperature {\bf(c)} - profile of electron potential energy, the band diagram {\bf(d)} - profile of local Seebeck coefficient. {\bf(e)} gate-dependent photovoltage computed with microscopic quantities (a-d) for various angles of linear polarization marked with colors. Inset in (e) shows the polarization ratio defined as ${\rm PR}(\theta) = V_{\rm ph}(\theta)/V_{\rm ph}(\theta=0)$. Dashed lines show the portion of photovoltage generated in the bulk of graphene {\bf(f)} polarization reconstruction quality $Q$ computed from the photovoltage in (e), $\log_{10}$ scale}
    \label{fig3}
\end{figure*}

Our model assumes the dominant photo-thermoelectric mechanism of photovoltage generation. This is concordant with prior experimental studies~\cite{Tielrooij-JPCM,Tielrooij-NNano,Cai_sensitive_THz} and justified theoretically by the relatively long electron cooling times $\tau_\varepsilon \sim 1$ ps~\cite{Levitov_PTE_PVE}~\footnote{The electron-hole photovoltaic effect is subleading due to the rapid Auger recombination $\tau_R \sim 100$ fs. The ratio of photon drag photovoltage to the thermoelectric voltage is $(\omega\tau_\varepsilon)^{-1}$, which is well below unity for infrared light.}  The photo-thermoelectric voltage appears as the convolution of position-dependent Seebeck coefficient $S(x,V_{\rm g})$ and local gradient of electron temperature,  $V_{\rm ph} = -\int_{0}^{\infty}{dx S(x,V_{\rm g}) T_e'(x)}$. To capture the entangled gate- and polarization-dependent photoresponse, it is essential to consider both realistic dependences of $S(x)$ and $T_e(x)$. This makes our model different from prior ones, where either simplified step-like profiles of $S(x)$ were adopted~\cite{Badioli2014,Optimizing_PTE}, or polarization-dependent near field modifications of local absorption were ignored~\cite{Cai_sensitive_THz,Xia_photocurrent_imaging}.

The model starts with evaluation of local infrared absorbance $\alpha(x)$ taking into account polarization-dependent near-field effects (see Methods and~\cite{Nikulin2021,Shabanov_Exact}). The absorbance $\alpha(x)$ peaks near the junction if the electric field ${\bf E}_0$ is orthogonal to the metal contact [red curve in Fig.~\ref{fig3} (a)], and drops down due to the skin-effect in metal if ${\bf E}_0$ is parallel to the junction [blue curve in Fig.~\ref{fig3} (a)]. Given the absorbed radiation power, we find the electron heating $\delta T_e = T_e - T_0$ by solving the heat conduction equation and assuming electron thermalization in metal contacts. The resulting heating profiles are shown in Fig.~\ref{fig3} (b). For $x$-polarized light, the carrier temperature reaches a maximum in the vicinity of the contact. For $y$-polarized light the temperature steadily grows to the constant value dictated by the bulk absorbance. As the polarization is gradually rotated between these limiting cases, the relative role of the two maxima is varied.

In the last step, we evaluate the local electrostatic energy $U_e(x)$ and Fermi energy $E_F (x) = -U_e(x)$ by solving the Poisson's equation with non-linear screening~\cite{Khomyakov_screening} and fixed metal-graphene doping $E_F(x=0)=0.1$ eV. Known the Fermi energy, we evaluate $S(x)$ with the conventional kinetic theory~\cite{Das_Sarma_Impurity,DasSarma_thermopower}. Several non-trivial examples of the band diagrams and thermopower profiles are shown in Fig.~\ref{fig3} (c) and (d). One observes that the gate voltage controls not only the 'bulk' thermpopower, but also the width of the Schottky barrier. The barrier width can change from low values $W_{\rm SB} \approx 100$ nm (for small carrier density in the bulk) to moderately high values $W_{\rm SB} \approx 600$ nm for lathe bulk density. In the latter case, the barrier width fully covers the carrier hot spot formed near the metal for $x$-polarization. Eventually, we convolve $S(x)$ with local temperature gradient $T_e'(x)$ to obtain the photovoltage $V_{\rm ph}$.

The resulting gate- and polarization-dependent photovoltages $V_{\rm ph}(V_{\rm g},\theta)$ are shown in Fig.~\ref{fig3} (e). Despite they resemble the gate-dependent Seebeck coefficient of bulk graphene $S(V_{\rm g})$ in general outlines, there appear pronounced features important for polarization-resolving action. Namely, the polarization ratio ${\rm PR}=V_{\rm ph}(\theta=\pi/2)/V_{\rm ph}(\theta=0)$ varies strongly with the gate voltage. It takes both large positive values at $-$20 V $<V_{\rm g} < 0$ V, negative values at $V_{\rm g}\lesssim -20$ V, and nearly-unity value at $V_{\rm g}\approx 5$ V. Large positive values are anticipated from lightning-rod enhancement of absorbance, while near-unity and negative values are not. The latter appear only due to the non-trivial spatial dependence of $S(x)$, which can change sign at some point between the metal contact and the bulk of graphene. 

The reconstruction quality computed theoretically is shown in Fig.~\ref{fig3} (f), it reaches moderately high values $Q \sim 0.1$ when one of the gate voltages is chosen close to the polarization-insensitive point $V_{\rm g}^*\approx 5$ V.

It is instructive to note that the conventional model of thermoelectric effect at a step-like metal-induced Schottky junction~\cite{Badioli2014,Optimizing_PTE} fails to predict the polarization-resolving response. 
The polarization-alternating sign of photovoltage does not appear in such a model, as polarization rotation affects only the magnitude of a positive quantity $\delta T_e (x=W_{\rm SB})$. The computed gate dependences of photovoltage within such model are shown with dashed curves in Fig.~\ref{fig3} (e) and confirm this statement.

{\it Discussion of the results.} We have demonstrated the feasibility of reconstructive mid-infrared polarimetry with detectors based on gate-tunable metal-graphene junctions. The conceptual observation enabling the polarization reconstruction lies in the 'entanglement' of gate- and polarization dependences of responsivity $R(V_{\rm g},\theta) \neq f(\theta) R(V_{\rm g})$. In other words, the gate voltage in the studied structures changes the polarization contrast in a controllable fashion, and readout of photoresponse in states with different polarization contrasts enables unambiguous determination of both power $P$ and linear polarization angle $\theta$. 

While we have not tested the feasibility of polarimetry in other ranges of electromagnetic spectrum, we may argue that the discovered functionality is specific to the infrared. The reason lies in commensurate scales of local absorption enhancement $\lambda_0/10 \sim 1$ $\mu$m and the width of photosensitive Schottky barrier $W_{\rm SB}\sim 0.1 ... 0.5$ $\mu$m. In the visible range~\cite{Echtermeyer2014}, the extent of hot carrier spot will be order of tens of nanomenters and, independent of gate voltage, will always fit into the Schottky barrier width. In terahertz range, on the contrary, all the 2d channels falls in the near-field domain of the contacts~\cite{Cai_sensitive_THz}. The polarization sensitivity of terahertz devices is governed by the geometry of their antennas and is not tunable by the gate voltage. Constraint of the polarimetry to the infrared part of spectrum still retains most important applications, such as chemical isomer identification and thermal imaging in under limited visibility.

Our model of entangled gate- and polarization-dependent response was based on the assumption of thermoelectric photovoltage signal. While there is strong theoretical evidence for the subleading role of photovoltaic~\cite{Levitov_PTE_PVE} and photon drag~\cite{Photon_drag} photocurrents, these mechanisms cannot be fully excluded. Their presence would not affect the functionality of reconstructive polarimetry. Indeed, the 'learning' equation \ref{eq-main_vec} is based only on general symmetry arguments, and its coefficients $R_{\rm is}$ and $R_{\rm an}$ are obtained with experimental calibration procedure.

Most of the devices of this study were based on scalable chemical vapor deposited graphene. The high-quality hBN-encapsulated samples, Fig.~\ref{fig2} (d), did not show any superior polarization reconstruction quality. The latter appeared to depend mainly on graphene-metal junction geometry. Indeed, the effects of local polarization-dependent field enhancement and suppression at the metal edges are insensitive to the conductivity of 2d material, and are rather governed by the thickness of metals. As these effects have purely electrodynamic origin, we can suggest a similar polarization-resolving functionality in other infrared-sensitive 2D electron systems, including palladium diselenide~\cite{PdSe2_detector}, black phosphorous~\cite{BlackP_detector} and mercury cadmium telluride quantum wells~\cite{HgTe_photonics}.

\section*{Methods}
\subsection*{Device fabrication}
The graphene detectors were fabricated on a B-doped Si substrate (resistivity 12 $\Omega\cdot$cm) with a 285~nm SiO$_2$ layer. At the first stage, CVD graphene was transferred from copper foil onto the substrate using a standard wet-transfer method with a PMMA supporting layer. Devices based on encapsulated graphene structures of the SiO$_2$/hBN/graphene/hBN type were fabricated using standard dry-transfer methods, employing a PC/PDMS stamp for assembly.

For both device types, a double-layer PMMA resist was deposited and patterned by electron-beam lithography to define the metal electrodes. Ti (2~nm)/Au (60~nm) metal contacts were deposited by electron-beam evaporation. At the final step, lithography was repeated, followed by plasma etching in an ICP-RIE system using O$_2$ gas for graphene and SF$_6$ gas for hBN, in order to define the channel geometry.

\subsection*{Optoelectronic measurements}
Optical measurements were performed in evacuated chamber with residual gas pressure $P\sim 10^{-4}$ Torr at room temperature. The radiation was fed from quantum cascade laser (QCL) with central wavelength $\lambda_0=8.6$ $\mu$m. Laser power was variable in the range $P=1...10$ mW by tuning the QCL drive current amplitude. The drive current was further modulated in a step-wise fashion with frequency $f_{\rm mod} = 911$ Hz, which enabled lock-in measurements of the photovoltage. A $\lambda_0/4$-waveplate with axes rotated by $45^\circ$ to initial polarization direction of laser radiation and a polarizer were introduced between laser and sample. This enabled power-preserving polarization rotation in our experiment. Rotation of polarizer was achieved through a programmable electro-mechanical driver.  

\subsection*{Modeling}
The local infrared absorbance $\alpha(x)$ by graphene for the two orthogonal light polarizations is defined as the local electromagnetic dissipation $q(x) = {\rm Re}\sigma(\omega)|{\bf E}_\omega(x)|^2/2$, where ${\rm Re}\sigma(\omega)\approx e^2/(4\hbar)$ is the universal optical conductivity of graphene, normalized by the intensity of the incident light $E_0^2/(2Z_0)$, where $Z_0$ us the free-space impedance. Importantly, the local field ${\bf E}_\omega(x)$ is different from the incident one ${\bf E}_0$, and strongly depends on position $x$ and polarization due to the lightning-rod effect at a metal contact. The dependence ${\bf E}_\omega(x)$ is computed analytically using the known solution for diffraction at a two-dimensional junction~\cite{Nikulin2021,Shabanov_Exact}. 

The carrier temperature $\delta T_e(x) = T_e(x)-T_0$ is computed at the second modeling stage by numerically solving the electron heat conduction equation:
\begin{equation}
   - \frac{\partial }{\partial x}\left[\chi(x)\frac{\partial \delta T_e}{\partial x}\right] =- \frac{C}{\tau_\varepsilon}\delta T_e(x) + q(x),
\end{equation}
where $\chi(x)$ is the carrier thermal conductivity, $C$ is their heat capacitance, $\tau_\varepsilon$ is the carrier cooling time, and $q(x) = {\rm Re}\sigma_{\rm opt}(w)|E_\omega(x)|^2/2$ is the local absorbed power. Complete electron thermalization at metal leads is assumed, $\delta T_e(x=0) = 0$. While the real electron thermal conductivity $\chi(x)$ is position-dependent, we observed only a minor effect of thermal conduction non-uniformity on $\delta T_e(x)$. Assuming constant thermal conductivity throughout the graphene $\chi={\rm const}$, we can write down the semi-analytical solution for temperature:
\begin{gather}
    \delta T_e(x) = \frac{\tau_\varepsilon}{C} \int_0^{+\infty}{dx'g(x,x')q(x')},\\
    g(x,x')=\frac{1}{L_\varepsilon}[e^{-|x-x'|/L_\varepsilon}-e^{-|x+x'|/L_\varepsilon}],
\end{gather}
where $g(x,x')$ is the fundamental solution of one-dimensional heat conduction equation, and $L_\varepsilon = \sqrt{\chi \tau_\varepsilon/C_e}$ is the energy relaxation length.

The local electrostatic energy  $U_e(x) = -e\varphi(x)$, where $\varphi(x)$ is the electric potential, and local Fermi energy $E_F(x) = -U_e(x)$, are found from the non-linear Poisson's equation
\begin{multline}
\label{eq-screening}
    \varphi(x) = V_{\rm g} {\mathcal F}_g(x) + V_s {\mathcal F}_s(x) + \\
    e \int_0^{+\infty}G(x,x')[p(x')-n(x')]dx'.
\end{multline}
Above, $V_{\rm g}$ and $V_s$ are the fixed values of electric potential at the gate and source leads, ${\mathcal F}_g(x)$ and ${\mathcal F}_s(x)$ are the 'characteristic potentials', i.e. spatial distributions of electric potential for unity voltage at gate (source) and all other leads grounded, $G(x,x')$ is the Green's function of electrostatic problem, $p(x)$ and $n(x)$ are the local densities of holes and electrons. While $V_{\rm g}$ is controlled in situ, $V_s$ is fixed by the metal-graphene work function difference. We adopt $V_s = 0.1$ V for graphene-gold contact.

Once $E_F(x)$ is found, the position-dependent Seebeck coefficient is computed from the known differential thermopower $a$ and electric conductivity $\sigma$
\begin{equation}
\label{eq-seebeck}
S = \frac{a(E_F)-a(-E_F)}{\sigma(E_F)+\sigma(E_F)},
\end{equation}
where the anti-symmetrization of $a$ and symmetrization of $\sigma$ are made to account for both electrons and holes. In turn, the electronic coefficients $a(E_F)$ and $\sigma(E_F)$ are found from the kinetic theory of electron-impurity scattering:
\begin{gather}
\label{eq-alpha}
a(E_F) = \frac{e}{2k} \int_{0}^\infty dE\, D(E) \frac{E - E_F}{kT} \frac{\partial f_0}{\partial E},  \\
\label{eq-sigma}
\sigma(E_F) = -\frac{e^2}{2} \int_{0}^\infty dE \, D(E) \frac{\partial f_0}{\partial E}.
\end{gather}
where $D(E)=\rho(E) v^2(E) \tau_p(E)$ is the energy-dependent diffusivity, $\tau_p(E)$ is the transport relaxation time, $v(E)$ is the group velocity of charge carrier at energy $E$ (reducing to a constant Fermi velocity $v_0\approx 10^6$ m/s in graphene), $\rho(E)=2 E / \hbar^2 v_0^2 \pi $ is the density of states, and $f_0$ is the equilibrium Fermi function. The 


\subsection*{Acknowledgements}
This work was supported by the grant No. 24-79-10081 of the Russian Science Foundation. Device fabrication was performed using the equipment of the Center of Shared Research Facilities (Moscow Institute of Physics and Technology).

\bibliography{references}

\end{document}